\documentclass[aps,pra,reprint,superscriptaddress]{revtex4-1}

\usepackage{amssymb}		\usepackage{graphicx}		\usepackage{xspace}
\usepackage{color}

\begin{document}

\let\oldSection\section

\renewcommand{\section}[1]{
\vspace{-11pt}
\oldSection{#1}
\vspace{-5pt}
}

\newcommand{\insertFigure}[4]{
\begin{figure}[t]
	\begin{center}
  	\includegraphics[width=#3]{#1}
  	\end{center}
  	\vspace{-1 em}
    \caption[#2]{#4}
\label{#2}
\end{figure}
}

\newcommand{\insertTwoImageFigure}[5]{
\begin{figure}[htbp]
	\begin{center}
  	\includegraphics[width=#4]{#1}
  	\includegraphics[width=#4]{#2}
  	\end{center}
    \caption[#3]{#5}
\label{#3}
\end{figure}
}

\newcommand{\figRef}[1]{Fig. \ref{#1}}

\newcommand*{\firstChange}[1]{#1}
\newcommand*{\secondChange}[1]{#1}
\newcommand*{\omittedChange}[1]{}

\newcommand{\snark}[1]{} 

\title{Student reasoning about measurement uncertainty in an introductory lab course}

\author{H. J. Lewandowski}
\affiliation{Department of Physics, University of Colorado Boulder, Boulder, CO 80309, USA} 
\affiliation{JILA, National Institute of Standards and Technology and University of Colorado Boulder, Boulder, CO 80309, USA}
\author{Robert Hobbs}
\affiliation{Department of Physics, University of Colorado Boulder, Boulder, CO 80309, USA} 
\affiliation{Department of Physics, Bellevue College, Bellevue, WA 98007, USA}
\author{Jacob T. Stanley}
\author{Dimitri R. Dounas-Frazer} 
\author{Benjamin Pollard}
\affiliation{Department of Physics, University of Colorado Boulder, Boulder, CO 80309, USA}

\begin{abstract}
Proficiency with calculating, reporting, and understanding measurement uncertainty is a nationally recognized learning outcome for undergraduate physics lab courses.
The Physics Measurement Questionnaire (PMQ) is a research-based assessment tool that measures such understanding. 
The PMQ was designed to characterize student reasoning into point or set paradigms, where the set paradigm is more aligned with expert reasoning. 
We analyzed over 500 student open-ended responses collected at the beginning and the end of a traditional introductory lab course at the University of Colorado Boulder. 
We discuss changes in students' understanding over a semester by analyzing pre-post shifts 
in student responses regarding data collection, data analysis, and data comparison.
\end{abstract}


\maketitle

\section{Introduction and Motivation}
Undergraduate physics lab courses provide students with opportunities to develop and practice experimental physics skills that are essential for many professional researchers. Such skills include data analysis and visualization, as well as the ability to compute and understand uncertainties~\cite{kozminski2014aapt}. Indeed, there is a national need for college graduates with these and other skills in a wide variety of physics and engineering career pathways~\cite{Heron2016,olson2012engage}. However, compared to lecture-based instruction, there is relatively little research on lab courses~\cite{national2012discipline}, and additional work is needed to identify effective teaching practices in these environments. In this paper, we describe an important aspect of a research-based transformation of PHYS 1140, a large-enrollment stand-alone introductory lab course at the University of Colorado Boulder (CU). In particular, we focus on establishing baseline data for pre-post shifts in students' understanding of measurement uncertainty, a learning outcome that has been identified as important by the faculty involved in the transformation process.

In our transformation effort, we rely on point and set paradigms to characterize student understanding of measurement uncertainty. 
Developed by researchers at the University of Cape Town and the University of York~\cite{Buffler2001, Campbell2005}, these paradigms describe two opposing ideas about uncertainty. 
A student using point-like reasoning believes that a single measurement can yield the ``true value" of a physical quantity, while a student using set-like reasoning recognizes that no individual measurement yields the true value. 
Point-like reasoners consider individual measurements independently of each other, while set-like reasoners consider 
a distribution of measurements with an associated mean and spread
~\cite{Buffler2009}. 
One of the main goals for the transformed lab course is to support students in developing a set-like understanding of measurement uncertainty, which is associated with expert-like understanding.

We use the Physics Measurement Questionnaire (PMQ)~\cite{Buffler2001, Volkwyn2008a} to classify student reasoning into the point or set paradigm. The PMQ is a free-response assessment that consists of several questions, or ``probes." In this study, we restrict our analysis to four probes: \textit{repeating measurements} (RD), \textit{using repeated measurements} (UR), \textit{same mean with different spread} (SMDS), and \textit{different mean with the same spread} (DMSS). The RD and UR probes were designed to measure students' understanding of measurement uncertainty in the context of data collection and data analysis, respectively. The SMDS and DMSS probes both focus on data comparison. The PMQ is typically administered before and after instruction, and it can be used to measure shifts in student understanding from one paradigm to another.

In this article, we use the PMQ to provide insight into the effectiveness of PHYS 1140 prior to transformation, i.e., as it has been traditionally taught for many years. Accordingly, we explore two related research questions: (i) Which of the four PMQ probes shows significant shifts in point- or set-like reasoning in our traditional lab course? and (ii) Which aspects of measurement uncertainty probed by the PMQ does our traditional lab course teach well, and which need to be improved?

\section{Background}
The PMQ was first implemented at the University of Cape Town~\cite{Buffler2001,Campbell2005,Volkwyn2008a}. 
It has since been used at Uppsala University~\cite{LippmannKung2006}, the University of Maryland College Park~\cite{Lippmann2003,Kung2005}, and North Carolina State University~\cite{Abbott2003}. 
In all of these previous studies, the PMQ has been implemented in introductory lab courses, and has been used in an evaluative capacity for traditional or transformed labs. 
We use the PMQ in a similar capacity in this study.

The RD, SMDS, and DMSS probes each have four parts: (i) a prompt that describes a hypothetical scenario from a lab activity, (ii) two or three statements about the scenario, (iii) a multiple-choice question asking students to select which statement they agree with most, and (iv) an open-response text box that students must use to explain their reasoning. 
The RD probe describes a single measurement, and students must explain whether (and how many) additional measurements are needed. 
The SMDS and DMSS probes both present two tables of five measurements. 
For the SMDS probe, the two data sets have the same mean but different spreads, and students must explain whether or not one data set is better than the other. 
For the DMSS probe, the two data sets have different means but the same spread, and students must explain whether or not the data sets are in agreement with one another. 
The UR probe has a slightly different format than the other probes: it presents one table of five measurements, and asks students to record a single value as the final result. 
Students must then explain their choice in an open-response text box. 
More information about PMQ probes can be found in Ref.~\cite{Volkwyn2008a}.

For all PMQ probes, students' written explanations are the primary data that are analyzed. 
Qualitative approaches, such as those described in the following section, allow researchers to determine whether students' explanations align with the point or set paradigms. 
Statistical methods can then be used to look for significant differences in students' reasoning from pre- to post-instruction. 
This mixed methods approach is more time consuming than strictly multiple-choice assessments (e.g., the Concise Data Processing Assessment~\cite{Day2011,Holmes2014,Day2016}), but also more informative: students' written explanations can help researchers understand \textit{how} and \textit{why} students' reasoning is shifting (or not shifting) on a particular probe.

\section{Context and methods}

Participants in this study were students who completed PHYS 1140 during Fall 2016. 
Students typically enroll in PHYS 1140 during their second semester of study, after completing an introductory mechanics course and while concurrently enrolled in an electricity and magnetism course. 
PHYS 1140 convenes twice per week, once each for lab and lecture. 
In the lab portion, students complete six two-week-long activities that focus on classical physics phenomena. 
Students are responsible for completing pre-lab questions, writing lab reports, completing homework assignments on error analysis, and participating in clicker questions during lecture. There are no exams.

Of the 588 students in our study, 8\% were physics majors, 63\% were engineering majors (excluding engineering physics), 18\% were non-physics science and math majors, and 11\% were majors in other disciplines. 
Most students who complete PHYS 1140 are white and/or male. 
For example, between 2004 and 2014, 76\% of the students who completed the course were men and 24\% were women; 74\% were white, 9\% were Asian American, 8\% were members of an underrepresented racial or ethnic group in the US, and 4\% were international students~\footnote{The CU Office of Planning, Budget, and Analysis, most recent available data.}.

We administered the PMQ to all PHYS 1400 students both before and after instruction. 
The PMQ was treated as an in-class assignment worth 2\% of the final course grade, graded for participation only. 
In this article, we focus only on data generated by the 525 students who completed both the pre-test and post-test.

We analyzed student responses to the PMQ using four coding schemes, one each for the RD, UR, SMDS, and DMSS probes. 
These schemes were based on codebooks initially developed by Volkwyn et al.~\cite{Volkwyn2005} in the context of their work at the University of Cape Town. 
To fully capture the range of responses in our dataset, we expanded these codebooks by adding additional codes. The research team collaboratively assigned each code in the codebooks a score of \textit{P}, \textit{S}, or \textit{N}, depending on whether the code represented point-like reasoning, set-like reasoning, or reasoning that did not fit into either paradigm. 
An \textit{N} was also assigned to responses that were off-topic, difficult to interpret, or had elements of both point- and set-like reasoning.

All responses were coded by one author (R.H.). 
Another author (B.P.) independently coded a random subset of 10\% of the pre-test responses to ensure reliability of the expanded codebook. 
To determine the inter-rater reliability of the process of assigning \textit{P}, \textit{S}, or \textit{N} codes to student responses, we computed the percent agreement (78\%) and Cohen's kappa statistic (0.63). We conclude that there was “substantial agreement” between the two raters~\cite{Cohen1960,Blackman2000}.

We looked for differences between pre- and post-test PMQ scores for each probe. In our analyses, we represented pre- and post-test score distributions in three ways: first, using all three possible scores (\textit{P}, \textit{S}, or \textit{N}) per student per probe; second, using \textit{P} or not-\textit{P} scores; and third, using \textit{S} or not-\textit{S} scores. When comparing pre-post distributions, we computed measures for both statistically and practically significant differences.
To determine if there were statistically significant differences between pre- and post-test distributions, we compared the distributions for each probe using the nonparametric Mann-Whitney U-test at the 5\% significance level~\cite{Mann1947}. To determine if there were practically significant differences, we computed the difference in the mean number of \textit{P} scores before and after instruction, and we compared those differences to the 95\% confidence interval of the number of pre-test \textit{P} scores. We used the variance of the multinomial distribution to determine 95\% confidence intervals. Means and confidence intervals were normalized by the total number of students in the population. We repeated this process for \textit{S} scores.

\insertFigure{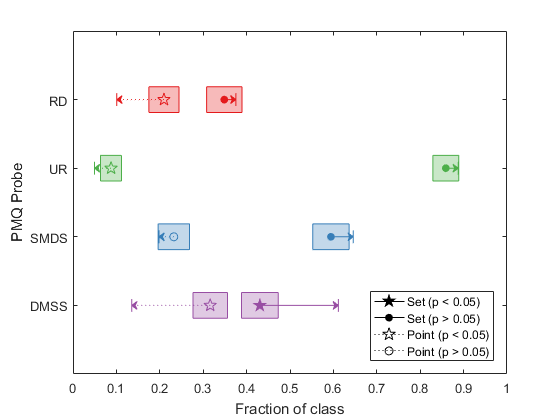}{ShiftsByQuestion}{\linewidth}{
Fraction of students with \textit{S} or \textit{P} (filled or open markers and dashed arrows, respectively) on each of four probes from the PMQ. 
Markers represent pre-test means. Ends of arrows represent post-test means.
Shaded boxes represent the 95\% confidence interval of pre-test values. 
The $p$ values in the legend refer to the result of a Mann-Whitney U-test of statistical significance between pre- and post-test distributions, with stars representing significant shifts and circles representing no significant differences at the 5\% significance level.
}
\section{Results and Discussion}
Figure \ref{ShiftsByQuestion} shows the fraction of students that had a \textit{P} or \textit{S} response on each of the four questions from the PMQ.
Solid and open markers show, respectively, the \textit{S} and \textit{P} fraction for the pre-test data.
The corresponding arrows show the same fraction for the post-test data.
The shaded boxes represent the the 95\% confidence interval of the pre-test data as a measure of practical significance.

We see a statistically significant difference across all probes between pre-test and post-test distributions of all paradigms taken together ($p \ll 0.01$ for RD and DMSS, $p = 0.01$ for UR, and $p = 0.03$ for SMDS).
We also see a statistically significant pre-post difference in whether responses were coded as \textit{P} in all probes except SMDS ($p \ll 0.01$ for RD and DMSS, $p = 0.01$ for UR), whereas the pre-post difference in just the \textit{P} paradigm was not significant for the SMDS probe ($p = 0.18$).
Likewise, for whether the responses were coded as \textit{S}, we see a statistically significant difference for DMSS ($p \ll 0.01$), but not for RD ($p = 0.37$), UR ($p = 0.16$), and SMDS ($p = 0.09$).
These findings are represented by markers in Fig. \ref{ShiftsByQuestion}, with stars representing statistically significant shifts.

The first probe, RD, shows low fractions of both \textit{S} and \textit{P} codes in both the pre-test and post-test, with the plurality of responses coded as \textit{N}.
Furthermore, the significant decrease in \textit{P} responses after instruction was not accompanied by a significant increase in \textit{S} responses, meaning that a larger fraction of student responses fell into the \textit{N} paradigm after instruction than before.

We consider two possible interpretations that could be drawn from this finding, each with different implications for course transformation.
One interpretation is that students were unable to articulate a consistent reasoning in the context of data collection at the start of the class, and that their inability to do so increased (on average) after instruction.
This interpretation suggests that our traditional class is failing to shift students towards more expert-like reasoning about measurement uncertainty in data collection.

An alternative interpretation is that the RD probe is not effectively prompting many students to use reasoning aligned with the point-set paradigms. 
This interpretation is supported by findings at Uppsala University, where this probe was shown ``not to distinguish among students who are experienced in the student laboratory at the secondary and introductory university level~\cite{LippmannKung2006}."
The RD probe is the only PMQ probe that deals with a single data point (as opposed to sets of two or more measurements).
Therefore, it is possible that this probe elicits a wider scope of ideas than the other probes, readily prompting reasoning
which is outside point-set paradigms.
Further and more qualitative analysis is needed on responses to this probe, in particular on the responses coded as \textit{N}, to better understand 
student reasoning
before and after instruction.

In contrast, the UR probe shows a high fraction of \textit{S} codes and a low fraction of \textit{P} codes both before and after instruction, showing that our students come into our class consistently using expert-like reasoning for data analysis, and that they continue to do so at the end of the course.
This result suggests that the UR probe provides low signal in our course context, and does not provide as meaningful a measure of changes in students' understanding of measurement uncertainty compared to the other probes.

The final two probes (SMDS and DMSS, which concern students' ability to compare data sets) show 
the largest shifts towards set-like reasoning, though only DMSS shows a statistically significant increase in \textit{S} responses after instruction.
Overall, these increases represent larger shifts towards expert-like reasoning when comparing data sets than when reasoning about collecting or analyzing data.
Furthermore, while DMSS starts with a smaller fraction of \textit{S} and a larger fraction of \textit{P} compared to SMDS, it also measures a significant decrease in \textit{P} after instruction while SMDS does not.
The shifts for DMSS are markedly larger than the other three probes, bringing the DMSS post-test fractions of \textit{P} and \textit{S} to roughly the same levels as the SMDS probe after instruction.

According to Buffler et al.~\cite{Buffler2001}, these two probes differ in how the issue of spreads is presented.
The SMDS probe (which concerns two data sets with the same mean but different spreads) ``presented the issue of spread explicitly," requiring students to recognize spread ``as a descriptor of the quality of a set of measurements~\cite{Buffler2001}."
In contrast, for the DMSS probe (which concerns two data sets with different means but the same spread), ``the notion of spread \ldots\ needed to be conceptualized and applied~\cite{Buffler2001}," suggesting that DMSS treats spread less explicitly than SMDS. 
In this light, our data suggest that students treat spread 
using set-like reasoning when confronted with it explicitly, even before instruction, but they are initially less set-like in their ability to use notions of spread implicitly.
However, after instruction in our traditional course, students consistently apply concepts related to spread whether those concepts are contextually explicit or not.

\section{Conclusion and Outlook}
These results answer the research questions posed above.
With regard to question (i), which concerns the probes themselves, each of the four probes we studied shows statistically significant pre-post shifts in all paradigms of understanding taken together, and all show nonzero shifts towards expert-like reasoning.
However, the UR probe shows a high fraction of set-like reasoning and a low fraction of point-like reasoning before instruction, providing little room for further improvement.
Correspondingly, shifts in the UR probe are of only marginal practical significance, or not at all.
In contrast, RD and DMSS each show practically and statistically significant shifts in either point or set paradigms, with DMSS showing the largest shifts towards \textit{S} and away from \textit{P}.

Question (ii) concerns the specific aspects of student understanding measured by our probes.
We see the largest shift towards more expert-like reasoning when students compare data sets, in particular when the notion of spread is contextualized only implicitly.
In contrast, while we see a significant decrease in point-like approaches after instruction when students reason about data collection, that shift was not accompanied by an increase in set-like reasoning.
Furthermore, data collection showed the lowest level of set-like reasoning overall, and also a low level of point-like reasoning, which could suggest that students both start and end unable to articulate their understanding about this aspect.
More analysis is needed to verify this finding due to the 
large fraction of responses to this probe coded as \textit{N}.

We note a limitation of 
point-set paradigms
in probing student understanding of measurement uncertainty.
The point-set paradigms encompass concepts surrounding statistical uncertainty, where the spread of a data set arises from random variations.
Systematic uncertainty, in which uncontrolled variations tend to skew data in a particular direction, is also an important aspect of measurement uncertainty, and is not treated by these paradigms.
Therefore, analysis of PMQ data based on point-set paradigms does not measure student understanding of systematic uncertainty, even if such information is contained in student responses.
Nonetheless, point-set analyses of PMQ data are established as a measure of student understanding of statistical uncertainty, and our results should be considered within that scope. 

Despite this limitation, these findings nevertheless provide guidelines for our ongoing transformation process of the introductory lab course at CU.
Our results suggest that instructors should highlight concepts of measurement uncertainty in the context of data collection. One way to do this would be to provide students with opportunities to make and justify decisions about whether or not to collect additional data for a given measurement.
The results presented here also support maintaining a focus on data comparison throughout the course.

Lastly, our ongoing work aims to more fully understand the variations contained within each paradigm classification.
More qualitative analysis is needed to better understand students' thought processes 
to better encompass the breadth of student understanding around measurement uncertainty.
Future work will also focus on characterizing the finer gradations within set- and point-like reasoning within each aspect probed by the PMQ.\\

\section{Acknowledgments}
Bethany Wilcox provided useful input on quantitative analysis methods. 
This material is based upon work supported by the NSF under Grant No. PHY-1125844, and by the University of Colorado.

\bibliography{references}

\end{document}